\documentclass[notitlepage,onecolumn,showpacs]{revtex4-1}
\usepackage[utf8]{inputenc}
\usepackage{amsmath}
\usepackage{amsfonts}
\usepackage{amssymb}
\usepackage{times,fullpage}
\usepackage{comment}
\usepackage{array,graphicx}
\usepackage{color}
\usepackage[colorlinks=true,citecolor=blue,linkcolor=blue,urlcolor=blue]{hyperref}%
\usepackage{subcaption}
\usepackage{tikz}
\usepackage{mathtools}
\usepackage{soul}
\usetikzlibrary{decorations.pathmorphing,patterns}

\graphicspath{{./figures/}}

\begin{document}

\newcommand{\nwc}{\newcommand}
\nwc{\vs}{\vspace}
\nwc{\hs}{\hspace}
\nwc{\la}{\langle}
\nwc{\ra}{\rangle}
\nwc{\nn}{\nonumber}
\nwc{\Ra}{\Rightarrow}
\nwc{\wt}{\widetilde}
\nwc{\lw}{\linewidth}
\nwc{\ft}{\frametitle}
\nwc{\ben}{\begin{enumerate}}
\nwc{\een}{\end{enumerate}}
\nwc{\bit}{\begin{itemize}}
\nwc{\eit}{\end{itemize}}
\nwc{\dg}{\dagger}
\nwc{\mA}{\mathcal A}
\nwc{\mD}{\mathcal D}
\nwc{\mB}{\mathcal B}

\nwc{\Tr}[1]{\underset{#1}{\mbox{Tr}}~}
\nwc{\pd}[2]{\frac{\partial #1}{\partial #2}}
\nwc{\ppd}[2]{\frac{\partial^2 #1}{\partial #2^2}}
\nwc{\fd}[2]{\frac{\delta #1}{\delta #2}}
\nwc{\pr}[2]{K(i_{#1},\alpha_{#1}|i_{#2},\alpha_{#2})}
\nwc{\av}[1]{\left< #1\right>}
\nwc{\alert}[1]{\textcolor{red}{#1}}

\nwc{\zprl}[3]{Phys. Rev. Lett. ~{\bf #1},~#2~(#3)}
\nwc{\zpre}[3]{Phys. Rev. E ~{\bf #1},~#2~(#3)}
\nwc{\zpra}[3]{Phys. Rev. A ~{\bf #1},~#2~(#3)}
\nwc{\zjsm}[3]{J. Stat. Mech. ~{\bf #1},~#2~(#3)}
\nwc{\zepjb}[3]{Eur. Phys. J. B ~{\bf #1},~#2~(#3)}
\nwc{\zrmp}[3]{Rev. Mod. Phys. ~{\bf #1},~#2~(#3)}
\nwc{\zepl}[3]{Europhys. Lett. ~{\bf #1},~#2~(#3)}
\nwc{\zjsp}[3]{J. Stat. Phys. ~{\bf #1},~#2~(#3)}
\nwc{\zptps}[3]{Prog. Theor. Phys. Suppl. ~{\bf #1},~#2~(#3)}
\nwc{\zpt}[3]{Physics Today ~{\bf #1},~#2~(#3)}
\nwc{\zap}[3]{Adv. Phys. ~{\bf #1},~#2~(#3)}
\nwc{\zjpcm}[3]{J. Phys. Condens. Matter ~{\bf #1},~#2~(#3)}
\nwc{\zjpa}[3]{J. Phys. A ~{\bf #1},~#2~(#3)}
\nwc{\zpjp}[3]{Pramana J. Phys. ~{\bf #1},~#2~(#3)}

\title{A stochastic heat engine using an active particle}
\author{Aradhana Kumari$^{1,a}$, P. S. Pal$^{2,b}$, Arnab Saha$^{3,c}$ and Sourabh Lahiri$^{1,d}$}

\affiliation{$^1$Department of Physics, Birla Institute of Technology, Mesra, Ranchi, Jharkhand 835215\\$^2$Department of Physics, University of Maryland, Baltimore County, Baltimore, MD 21250, USA\\ $^3$Department of Physics, Savitribai Phule Pune University, Ganeshkhind, Pune 411007, India.}

\email{$^a$aradhanakumari2546@gmail.com,\\$^b$pspal@umbc.edu,\\$^c$sahaarn@gmail.com,\\$^d$sourabhlahiri@gmail.com}

%\date{}

\begin{abstract}
The topic of microscopic heat engine has undergone intensive research in recent years. Microscopic heat engines can exploit thermal as well as active fluctuations to extract thermodynamic work. We investigate the properties of a microscopic Stirling's engine that uses an active (self-propelling) particle as a working substance, in contact with two thermal baths. It is shown that the presence of activity leads to an enhanced performance of the engine. The efficiency can be improved by increasing the activity strength for all cycle time, including the non-quasistatic regime. We verify that the analytical results agree very well with our simulations. The variation of efficiency with the temperature difference between the two thermal baths has also been explored. The optimum region of operation of the engine has been deduced, by using its efficient power as a quantifier. Finally,  a simple model is provided that emulates the behaviour of a flywheel driven by this engine.
\end{abstract}
\pacs{}
\maketitle
\section{Introduction}

Stochastic thermodynamics has been the cornerstone of nonequilibrium thermodynamics and statistical mechanics in the last two decades \cite{sekimoto,sek98,sei05_prl,sei08_epjb,sei12_rpp}. It prescribes definitions of thermodynamic quantities like work, heat or entropy for individual experiments performed on small systems (dimensions $\lesssim 1 \mu m$)  \cite{sekimoto,sek98}, where the dynamics involves appreciable amount of thermal fluctuations. They are well-defined even if the system is driven far from equilibrium.

Since driving machines at such small scales requires commensurate engines, a lot of effort has been invested in trying to understand the thermodynamics of small engines in the last decade.
A stochastic Carnot heat engine was conceived in \cite{sei08_epl}, where the engine was modelled as a Brownian particle trapped in a harmonic potential. The stiffness of the trap varied slowly with time, so as to mimic the isothermal expansion and compression arms of the Carnot cycle. On the other hand, the adiabatic expansion and compression were imposed by a sudden change in the stiffness of the potential, so that no heat transfer takes place in the process. 
Such engines have also been realized experimentally using colloidal particles or even single atoms \cite{rol16_nature,rob16_science}. An autonomous stochastic heat engine was investigated in another recent experiment \cite{gar16_prl}. For a recent review, see \cite{rol17_sm}.

A Stirling heat engine for a mesoscopic particle was experimentally realized in \cite{bec12_nature}. Here, a particle trapped in a harmonic potential was subjected to a quasi-static change in the stiffness of the confining trap, thus mimicking the isothermal arms of the Stirling engine. The isochoric arms were modelled by an instantaneous change in the temperature of the bath, without any change in stiffness constant. Our present study will be based on this model.

%\sethlcolor{cyan}

The underlying principle is simple: the first step is an isothermal expansion (implemented by a quasistatic decrease in trap stiffness). Second step involves a sudden decrease in temperature, thus leading to isochoric (constant-volume) heat release. The third and fourth steps are isothermal compression and isochoric heat absorption, respectively. 
We, however, relax the constraint of the quasistatic nature of the isothermal arms of the cycle, and compute the relevant thermodynamic quantities for this system. To out knowledge, the engine parameters  have not yet been investigated in a nonequilibrium steady state. The detailed process is discussed in the next section.

Recently, the properties of an engine formed of a passive colloidal particle (i.e. particles that are not self-propelling) in bacterial (active) baths have been explored in the experimental work in \cite{ajay2016_nature} and the theoretical work in \cite{sah19_jsm}. An active bath consists of self-propelling particles that may be living like bacteria, or artificially prepared like the Janus particles \cite{ram13_rmp,har16_rmp} (for a review on Janus particles, see \cite{zha17_lan} and the references therein).
Both these works conclude that the efficiency of the engine may be made to surpass the efficiency obtained in the case of thermal baths in a suitable range of parameters. The statistical properties of micron-sized beads in a bacterial bath was studied earlier in \cite{lib00_prl}.

We consider our system of interest to be an active Ornstein-Uhlenbeck particle (AOUP) \cite{man17_prl,fod16_prl}.
The particle is immersed in a thermal environment whose temperature can assume the values $T_h$ or $T_c$, where $T_h>T_c$. The activity of the particle and temperature of the bath are time-periodically modulated. For the first half of the cycle, the particle remains passive and in contact with the hotter thermal bath. In the second half, it becomes active and is placed in contact with the colder bath.
 The activity may be controlled, for instance, by means of photo-activation \cite{but12_jpcm,usp19_jcp}.
When the particle is active, the activity will induce a correlated noise to the dynamics of the particle, in addition to the thermal noise, which can influence the work extraction considerably (see sec. \ref{sec:work_stirling}). We provide analytical solutions for the transient case for the system without imposing any constraint on the speed at which the process is carried out (i.e. not taking the quasistatic limit). Further, in contrast to \cite{sah19_jsm}, we do not ignore the effects of thermal noise in comparison to the active noise of the particle. We explore the possibility of obtaining engines that are comparatively more efficient than their passive counterparts. A related work for a system consisting of an active and a passive particle on a lattice has been done recently \cite{sei19_prx}.

In section \ref{sec:model}, we explain the model and the governing equations for the dynamics. In sec. \ref{sec:efficiency_QS}, the expression for efficiency for a passive microscopic Stirling cycle that works as per this model has been provided. Sec. \ref{sec:work_stirling} outlines the steps to find the work output from an active microscopic Stirling engine. We discuss the efficiency of this engine in sec. \ref{sec:efficiency_nonQS}. In sec. \ref{sec:efficient_power}, the maximization of the so-called efficient power has been discussed. Sec. \ref{sec:application}, we provide a simple model that shows how the output work from the engine can be used by a second particle, thus emulating the action of a flywheel attached to an engine.

\section{The Model}
\label{sec:model}

We consider a system consisting of a single active colloidal particle placed in a harmonic trap. The particle is inside a medium whose temperature can be switched between two values: a higher temperature $T_h$ and a lower temperature $T_c$. They form the hot and the cold thermal reservoirs for the engine, respectively.

\begin{figure}[!ht]
 \centering
 \includegraphics[width=7.5 cm,height=5.5 cm]{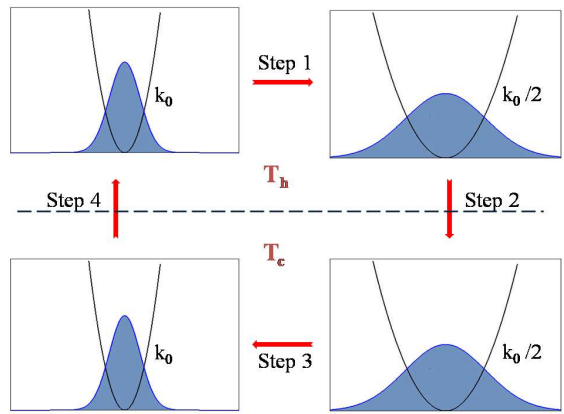}
 \caption{The Stirling cycle using optical tweezers. Step 1 is the isothermal expansion, step 2 is isochoric decrease in temperature, step 3 is isothermal compression and step 4 is isochoric increase of temperature. The spring constants of the confining potential have been mentioned in the figure.}
 \label{fig:stirling}
\end{figure}

The Brownian particle follows the overdamped Langevin equation. The Stirling cycle consists of two isothermal (constant temperature) arms and two isochoric (constant volume) arms, thus completing the four arms of the cycle (see figure \ref{fig:stirling}). 
The sequence is: isothermal expansion arm $\to$ isochoric arm (temperature is decreased in this step) $\to$ isothermal compression arm $\to$ isochoric arm (temperature is increased to the initial value in this step). In the context of the model of stochastic heat engine that we are dealing with, namely a colloidal particle in a harmonic trap formed by optical tweezers, in the isochoric arms of the cycle the stiffness of the optical trap remains constant. This is because the value of stiffness constant is inversely proportional to the volume available to the particle, so that an increase in stiffness constant would restrict the motion of the particle to a smaller volume.

The cycle is switched on at time $t=0$ and is completed at $t=\tau$. In the expansion process (which takes place for time $\tau/2$), the particle's position follows the Langevin equation
\begin{align}
  \hspace{1cm}\gamma\dot x &= -k_e(t)x + \big(\sqrt{D_h}\big)\xi.%, \hspace{0.3cm}0< t \le\frac{\tau}{2}~\mbox{(Expansion)}.
  \label{eq:lang_exp}
\end{align}
Here,  $D_h=2\gamma k_BT_h$, where $T_h$ is the temperature of the hot bath and $k_B$ is the Boltzmann constant, while $\xi$ is the thermal white noise that has zero mean and is Gaussian in nature. The time correlation of $\xi$ is $\av{\xi(t)\xi(t')} = \delta(t-t')$.
The time-dependence of the stiffness constant is given by
\[
  k_e(t) = k_0(1-t/\tau).
\]
This ensures that the stiffness of the trap is initially equal to $k_0$, and thereafter decreases (becomes flatter) as time increases, so that the final value is $k_0/2$;

In the second step, the temperature is suddenly changed from $T_h$ to $T_c$, with the value of stiffness constant held fixed at $k_0/2$.

In the third step, the system is held in contact with the cold bath and the stiffness constant is varied with the time-dependence given by
\[
  k_c(t) = k_0 t/\tau.
\]
This is the isothermal compression, in which the stiffness constant changes from $k_0/2$ to $k_0$, as the time changes from $\tau/2$ to $\tau$. In this step, the activity of the particle comes into effect. Thus, the Langevin equation now becomes
\begin{align}
  \hspace{1cm}\gamma\dot x &= -k_c(t)x + (\sqrt{D_c})\xi +\left(\sqrt{D_\eta/\tau_\eta}\right)\eta;\nn\\ 
  %\hspace{1cm}\frac{\tau}{2}\le t<\tau ~~~~\mbox{(Compression)}\nn\\
  \tau_\eta\dot\eta &= -\eta + \big(\sqrt{2\tau_\eta}\big)\xi_\eta.
                      \label{eq:lang_com}
\end{align}
Thus, in addition to the white noise of the Langevin equation in \eqref{eq:lang_exp}, we have a Gaussian distributed exponentially correlated (Ornstein-Uhlenbeck) noise \cite{ris,man17_prl,fod16_prl} (a recent work exploring the statistics of non-Gaussian active noise  is done in \cite{cha19_jcp}).  $\xi_\eta$ is another Gaussian white noise with zero mean. The initial distribution of $\eta$ is a delta-function at $\eta=0$. The correlation of $\eta$ is given by
\begin{align}
  \av{\eta(t)\eta(t')} = e^{-|t-t'|/\tau_\eta}.
  \label{eq:eta_corr}
\end{align}
The active noise strength $D_\eta$ is defined as in Eq. \eqref{eq:lang_com}. 
 $\tau_\eta$ is the correlation time for the active noise.

Finally, in the fourth step, the temperature is suddenly increased to $T_h$ and the activity is turned off, thus completing the full Stirling cycle.

\section{Efficiency of a passive quasistatically driven Stirling engine}
\label{sec:efficiency_QS}

Let the activity of the engine described above be set to zero (i.e. $D_\eta=0$ in Eq. \eqref{eq:lang_com}), so that the engine is now passive in nature. 
When the time taken to carry out the expansion and compression steps is very high (compared to the relaxation time to equilibrium) i.e. in the limit $\tau\to \infty$, these steps are quasistatic. However, it must be noted that this limit does not imply that the engine is reversible, since the sudden temperature changes in steps 3 and 4 make these steps irreversible. So the efficiency of a Stirling engine is always less than that of a reversible engine, i.e. the Carnot efficiency $\eta_c$. The expression is given by the relation (see appendix \ref{sec:appendix}, also \cite{bec12_nature})
\begin{align}
\eta_{stirling} &= \frac{\eta_{c}} {1+\eta_{c}\Big\{\ln\big[\frac{k_{max}}{k_{min}}\big]\Big\}^{-1}},
\label{quasistatic}
\end{align}
where $k_{max}$ and $k_{min}$ are the maximum and minimum values of the stiffness constant of the harmonic trap, which in our case are $k_0$ and $k_0/2$, respectively. The Carnot efficiency $\eta_c$ is given by $\eta_c=1-T_c/T_h$. It can be readily checked that for $T_h\gg T_c$, we have $\eta_c\to 1$, and the efficiency becomes independent of the temperature difference between the thermal reservoirs. In section \ref{sec:efficiency_nonQS}, we will find that the same qualitative behaviour is observed even when the Stirling engine is driven non-quasistatically in presence of active noise.

\section{Work extracted from an active Stirling engine}
\label{sec:work_stirling}

We now introduce activity in our system. For convenience, we consider a single active particle, whose activity is negligible at the higher temperature (expansion step), while it is appreciable at the lower temperature (compression step). The system follows the Langevin equations given by \eqref{eq:lang_exp} and \eqref{eq:lang_com} in the expansion and compression steps, respectively. We provide the steps to reach at the analytical result for extracted work and compare them with our simulations, in the next two sections. 
All the plots are with respect to dimensionless variables, where we have normalized time by $\tau$ and position by $\sqrt{(D_h-D_c)\tau}$. Defining $D\equiv D_h-D_c$, this would yield the normalized variables: $\tilde\gamma = 1, ~~\tilde k(t) = k(t)\tau/\gamma, ~~\tilde\xi = \xi\sqrt\tau/\gamma,
  ~~\tilde D_h = D_h/D, ~~\tilde D_c = D_c/D, ~~\tilde\tau_\eta =\tau_\eta/\tau,~~
   \tilde\eta = \eta/(\gamma D), ~~\mbox{and}~ \tilde D_\eta = D_\eta.$
We remove the tilde symbols in the plots for convenience.

\subsection{Time-evolution of the variance in position}

We now proceed to solve for the variance in the position of the particle as a function of time. We start the process at time $t=0$, where the particle always begins from a given initial position $x(0)=0$. Note that this is the transient regime, in which the system has not yet settled into a time-periodic steady state.  We shall solve the full dynamics given by  \eqref{eq:lang_exp} and \eqref{eq:lang_com}. In the next subsection, we would use this expression for variance to evaluate the work done in the first cycle.
 
 The solution for $x(t)$ in the expansion and compression steps are respectively
\begin{widetext}
\begin{alignat}{2}
  x(t) &= x_0e^{-I_e(t)} + \frac{\sqrt D_h}{\gamma}e^{-I_e(t)}\int_0^t \xi(t')e^{I_e(t')}~dt' \hspace{5cm}&&\mbox{(Expansion)} \label{eq:x_exp_gen} \\
  x(t) &= x(\tau/2)e^{-I_c(t)} + \frac{e^{-I_c(t)}}{\gamma}\int_{\tau/2}^t \left[(\sqrt D_c) \xi(t') + \left(\sqrt{\frac{D_\eta}{\tau_\eta}}\right)\eta(t')\right]e^{I_c(t')}~dt'. &&\mbox{(Compression)}
  \label{eq:x_com_gen}
\end{alignat}
%\end{widetext}
Here, $I_e(t)=\int_0^t dt' k_e(t')/\gamma$ and $I_c(t)=\int_0^t dt' k_c(t')/\gamma$. Let the variances in the expansion and compression cycles be given by $\sigma_e(t)=\av{x^2(t)}|_{0<t\le\tau/2}$ and $\sigma_c(t)=\av{x^2(t)}|_{\tau/2<t\le \tau}$.  Correspondingly, the dynamical equations for the variance in position are:
%
%\begin{widetext}
\begin{alignat}{2}
  \gamma\dot\sigma_e &= -2k_e(t)\sigma_e + 2\sqrt{D_h} \av{\xi(t)x(t)}; \hspace{4cm}&&\mbox{(Expansion)} \label{eq:sigma_exp_gen}\\
  \gamma\dot\sigma_c &= -2k_c(t)\sigma_c + 2\sqrt{D_c}\av{\xi(t)x(t)}+ \sqrt{\frac{D_\eta}{\tau_\eta}} \av{\eta(t)x(t)}. &&\mbox{(Compression)}
  \label{eq:sigma_com_gen}
\end{alignat}
%\end{widetext}
%
In order to respect causality, we must have $\av{x(\tau/2)\xi(t)}=0=\av{x(\tau/2)\eta(t)}$ for $t>\tau/2$. We therefore obtain $\av{\xi(t)x(t)} = \sqrt D_h/2\gamma$ in the expansion step and $\sqrt{D_c}/2\gamma$ in the compression step, so that from Eq. \eqref{eq:x_com_gen} we get
%
%\begin{widetext}
\begin{align}
  \av{\eta(t)x(t)} &= \sqrt{\frac{\pi\tau D_\eta}{2k_0\gamma\tau_\eta}} ~ \exp\left[-\frac{\left(k_0 t \tau _\eta+\gamma  \tau \right)^2}{2 k_0 \gamma  \tau  \tau
                     _\eta^2}\right]
                     \left\{\text{erfi}\left(\frac{k_0 t \tau _\eta+\gamma  \tau }{\tau_\eta\sqrt{2k_0\gamma\tau} }\right)-\text{erfi}\left(\frac{\sqrt{\tau } \left(k_0 \tau _\eta+2 \gamma
   \right)}{2\tau_\eta \sqrt{2k_0\gamma}}\right)\right\}.
   \label{eq:general}
\end{align}
%\end{widetext}
%
Here, $\text{erfi}(z)\equiv -i ~\text{erf}(iz)$ is the imaginary error function. For the first cycle (transient regime), we choose the initial position of the particle to be fixed at the origin for convenience. We then arrive at the following expression for the time evolution of the variance in the expansion step $\sigma_e(t)$:
%
%\begin{widetext}
\begin{align}
  \sigma_e(t) &= \left(\frac{D_h}{2 \gamma ^{3/2}}\sqrt{\frac{\pi\tau}{k_0}}\right)e^{\frac{k_0 (t-\tau )^2}{\gamma  \tau }}
   \left[\text{erf}\left(\sqrt{\frac{k_0\tau}{\gamma}}\right)-\text{erf}\left(\sqrt{\frac{k_0}{\gamma\tau}}(\tau-t)\right)\right].
\end{align}
\end{widetext}
\begin{figure}
\centering
  \begin{subfigure}[t]{0.48\linewidth}
  %\centering
  \includegraphics[width=\linewidth]{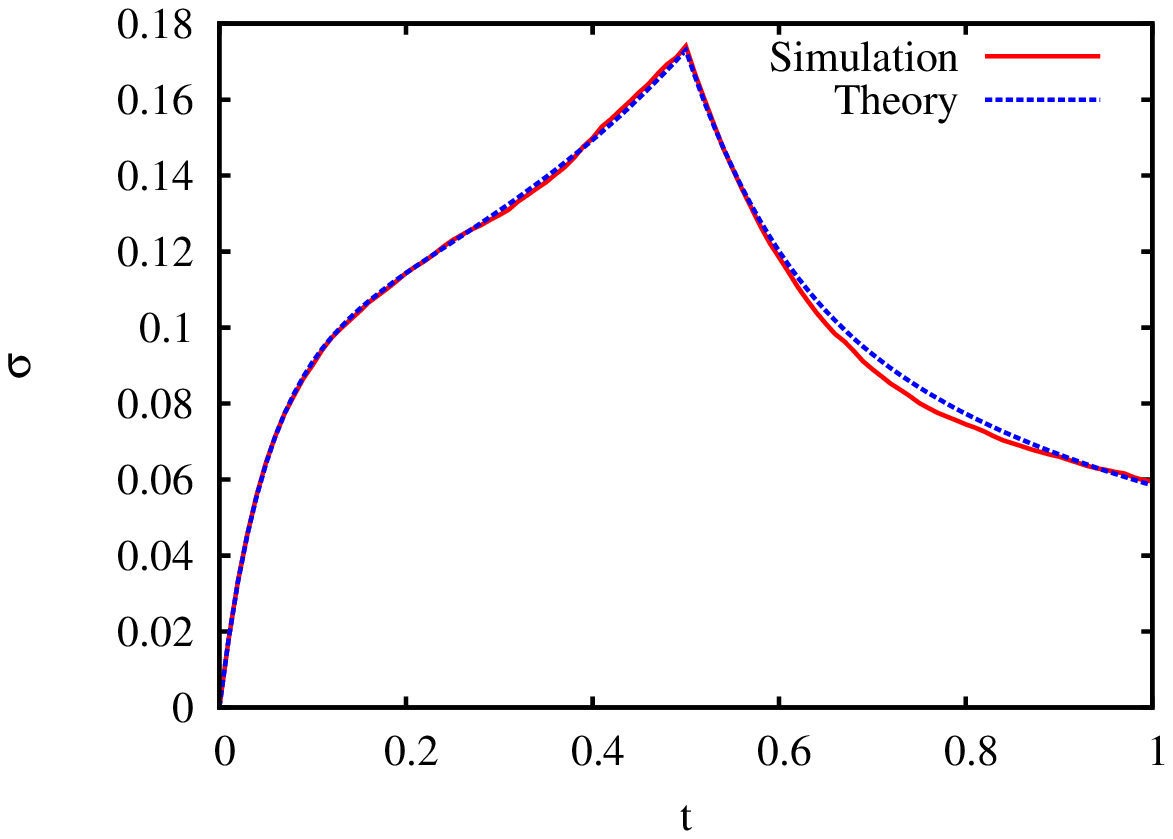}
  \caption{}
  \label{fig:sigma_cycle1}
  \end{subfigure}
  \hfill
  \hspace{-0.2cm}
  \begin{subfigure}[t]{0.48\linewidth}
  %\centering
 \includegraphics[width=\linewidth]{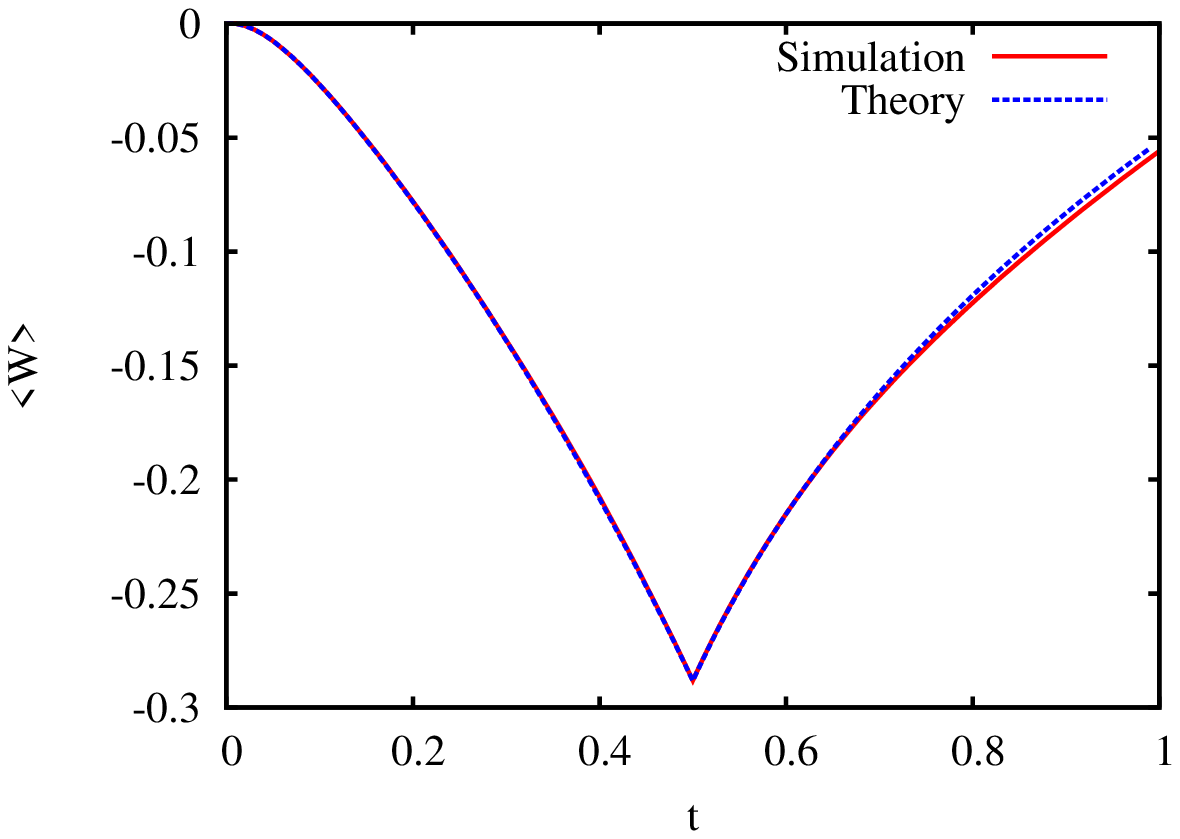}
 \caption{}
 \label{fig:work}
\end{subfigure}
\caption{(a) Plot of $\sigma(t)$ values when the two baths are at different noise stengths $D_h=2$ and $D_c=1$, in addition to an active noise in the compression cycle. Other parameters are: $\tau=1$ and $\tau_\eta=1$. (b) The time-variation of work done on the engine. The red line corresponds to results of simulation, while the blue line is the analytical result. The parameters are: $\tau=1$, $D_h=2$, $D_c=1$, $k_0=10$, $\tau_\eta=1$.}
\end{figure}
The expression for $\sigma_c(t)$ involves integrals that can be computed numerically (see appendix  \ref{sec:appendix2}).  We show a plot of the resulting solutions in figure \ref{fig:sigma_cycle1}.
The excellent agreement between the simulated and the analytical results even for such a highly nonlinear function of time provides a stringent test on the accuracy of our simulations.  For all simulations in this article, we have used the Heun's method to integrate the Langevin equations \cite{mannella}, and the averaging in each case has been done over $\sim 10^5$ trajectories.

\subsection{Calculation of mean work}

\begin{figure}[!ht]
 \centering
 \includegraphics[width=0.6\linewidth]{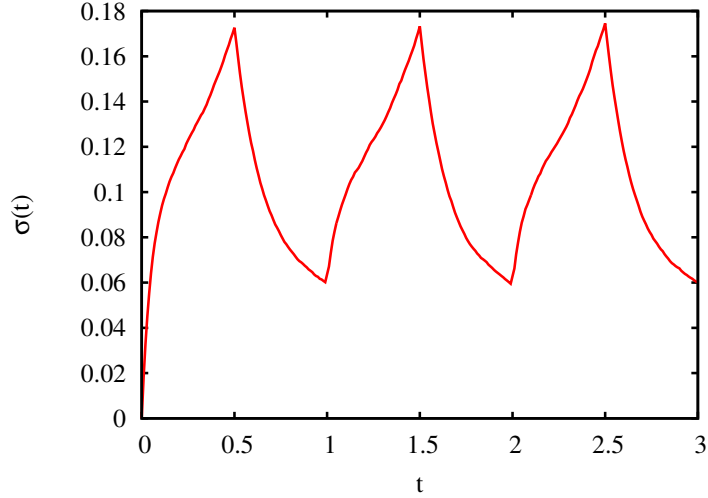}
 \caption{The time-variation of the variance for first three cycles. The parameters are: $\tau=1$, $D_h=2$, $D_c=1$, $k_0=10$, $\tau_\eta=1$.}
 \label{fig:TPSS}
\end{figure}

Once the variance of the system has been obtained as a function of time, the work can be calculated by using the definitions of stochastic thermodynamics \cite{sekimoto,sek98}
\begin{align}
 \left<W\right> &=  \frac{1}{2}\int_0^{\tau/2} \dot k_e(t) \sigma_{e}(t) dt + \frac{1}{2}\int_{\tau/2}^\tau \dot k_c(t) \sigma_{c}(t) dt.
 \label{W_gen}
\end{align}
Using the expressions for $\sigma_e(t)$ and $\sigma_c(t)$, the analytical expression for mean work can be obtained as a function of time for the first cycle in the transient regime. 
In figure \ref{fig:work}, we have plotted the time-dependence of work during the first cycle. The red solid line is the simulation result, while the blue dashed line gives the analytical result. 
We again find that the simulations agree with analytics with a high accuracy.
We also note from the figure that the qualitative features of the curves are same as for a passive particle, namely the extraction of work during the expansion step and expense of work during the compression step.
We now simulate the system in the steady state regime, where the initial distribution in no more a delta-function, but is a Gaussian distribution as shown in figure \ref{fig:stirling}. We observe from figure \ref{fig:TPSS} for the chosen set of parameters (mentioned in the figure caption), that by the time the system reaches the second cycle it has already reached a time-periodic steady state. To ensure that the system is always in the steady state for all the parameter ranges that we have used, the first three cycles are left out and the origin of time has been set to the beginning of the fourth cycle.

% \begin{figure}[!ht]
%   \centering
%   \includegraphics[width=0.6\linewidth]{engine_priyo_arnab.eps}
%   \caption{Comparison of $\sigma(t)$ values when the bath noise is switched off in the compression arm of the cycle in the general case for a single bath.}
% \end{figure}

\section{Calculation of efficiency of the active Stirling engine}
\label{sec:efficiency_nonQS}

The efficiency in presence of activity can be computed using the standard definition: $\eta=\left<W_{ext}\right>/\left<Q_{abs}\right>$, where the mean extracted work is $\av{W_{ext}} = -\left<W\right>$, while $\av{Q_{abs}} = -\left<Q_h\right>$ is the mean heat absorbed from the hot bath. 
In order to obtain $\left< Q_h\right>$, we use the First Law: $\av{Q_h} = \av W-\av{\Delta E} \equiv k_{min}\av{x^2(\tau/2)}/2 - k_{max}\av{x^2(0)}/2$. Here, $\Delta E$ is the change in energy during step 1 (isothermal expansion, see figure \ref{fig:stirling}), while $k_{max}=k_0$ and $k_{min}=k_0/2$ are the maximum and minimum values of the trap stiffness, respectively. 
We use this engine to explore the possibility of obtaining an engine that is more efficient than the usual Stirling engine that uses a passive particle as its working substance. 

\begin{figure}[!ht]
 \centering
 \includegraphics[width=0.6\linewidth]{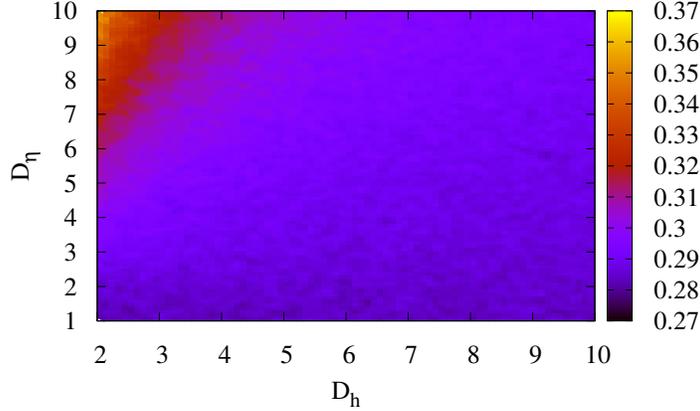}
 \caption{Phase plot of activity strength, thermal noise strength of hot bath and the efficiency of the engine. The colour coding provides the values of the efficiency. The parameters used are $\tau_\eta=1,~\tau=1$ and $D_c=0.1$.}
 \label{fig:Dh_Eta_efficiency}
\end{figure}

We first choose a set of parameters for which the system actually functions as an engine (it can also act as a refrigerator or as a dud \cite{pal14_pre}, which are of no concern for our present purposes). In figure \ref{fig:Dh_Eta_efficiency}, we find that the engine does so range of parameters used, so that we have $\av W<0$ (work is extracted) and $\av{Q_h}<0$ (heat is absorbed from the hot thermal bath).

\begin{figure}[!ht]
 \centering
 \includegraphics[width=0.6\linewidth]{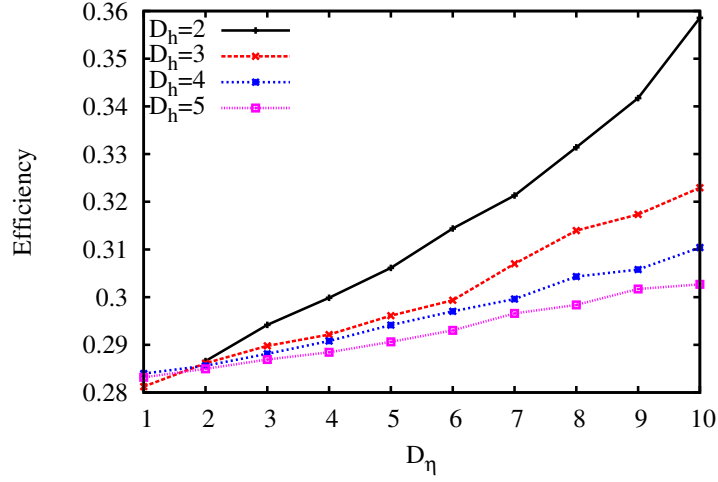}
 \caption{The plots show efficiency as a function of the bacterial activity. The increased efficiency for higher activity is apparent. The parameter values are $k_0=10,~\tau=1,~\tau_\eta=1,~D_c=0.1$.}
 \label{fig:efficiency_Deta}
\end{figure}

\begin{figure}[!ht]
 \centering
 \includegraphics[width=0.6\linewidth]{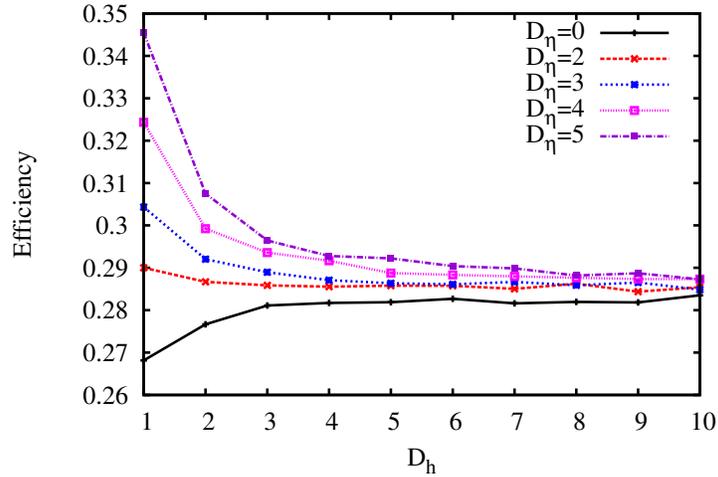}
 \caption{The plots show efficiency as a function of the noise strength of hot bath. The parameter values are $k_0=10,~\tau=1,\tau_\eta=1,~D_c=0.1$.}
 \label{fig:efficiency_Dh}
\end{figure}

In figure \ref{fig:efficiency_Deta}, we have shown the variation of the efficiency of our active Stirling engine as a function of the activity strength. We find that the efficiency increases with an increase in $D_\eta$, which means that a strongly active particle is in general more efficient than a weakly active one. Another interesting point to note is that for the same activity strength, efficiency decreases as the thermal noise of the hotter bath in increased. This is because a high temperature difference $\Delta T=T_h-T_c$ between the two thermal baths implies that it is the thermal force that is the dominant drive for the engine, and the contribution of active noise is much smaller. Thus, the engine tends towards a passive one with increase in $\Delta T$. Since the efficiency of a passive Stirling engine becomes independent of $\Delta T$ when $T_h\gg T_c$ (see figure \ref{fig:efficiency_Dh}), the efficiency becomes less as compared to an active Stirling engine.  

Figure \ref{fig:efficiency_Dh} shows the variation of efficiency with the thermal noise strength $D_h$ of the hot bath. We find that for the passive particle ($D_\eta=0$, black solid line), the efficiency increases with increase in $D_h$ and saturates at the value $\approx 0.28$.
It is observed that the efficiency of the passive Stirling engine becomes independent of temperature difference between the thermal reservoirs, just as in the quasistatic case (see sec. \ref{sec:efficiency_QS}).
For the corresponding curves for the active engine, the efficiency decreases with an increase in $D_h$, which is in accordance with our observations for figure \ref{fig:efficiency_Deta}. For a fixed value of $D_h$, we note that the efficiency increases with increase in $D_\eta$, again corroborating our observations for figure \ref{fig:efficiency_Deta}.

\section{Efficient power of the engine}
\label{sec:efficient_power}

In general, power and efficiency are not maximized simultaneously. In a reversible engine, the efficiency gets maximized, but the power drops to zero since the system is quasistatic. The Curzon-Ahlborn efficiency provides the approximate efficiency of an engine when it is working at its maximum power \cite{cur75_ajp}. 
It has been shown that there is a trade-off between output power of an engine and its efficiency, i.e. a higher power entails a lower efficiency \cite{sai16_prl,sei18_prl}. One of the ways in which these two parameters can be optimized is by maximizing the \emph{efficient power}, i.e. the product of its power and efficiency \cite{yil06_jei,joh18_pre}.
\begin{figure}[!ht]
\centering
\begin{subfigure}{0.48\linewidth}
 \includegraphics[width=\linewidth]{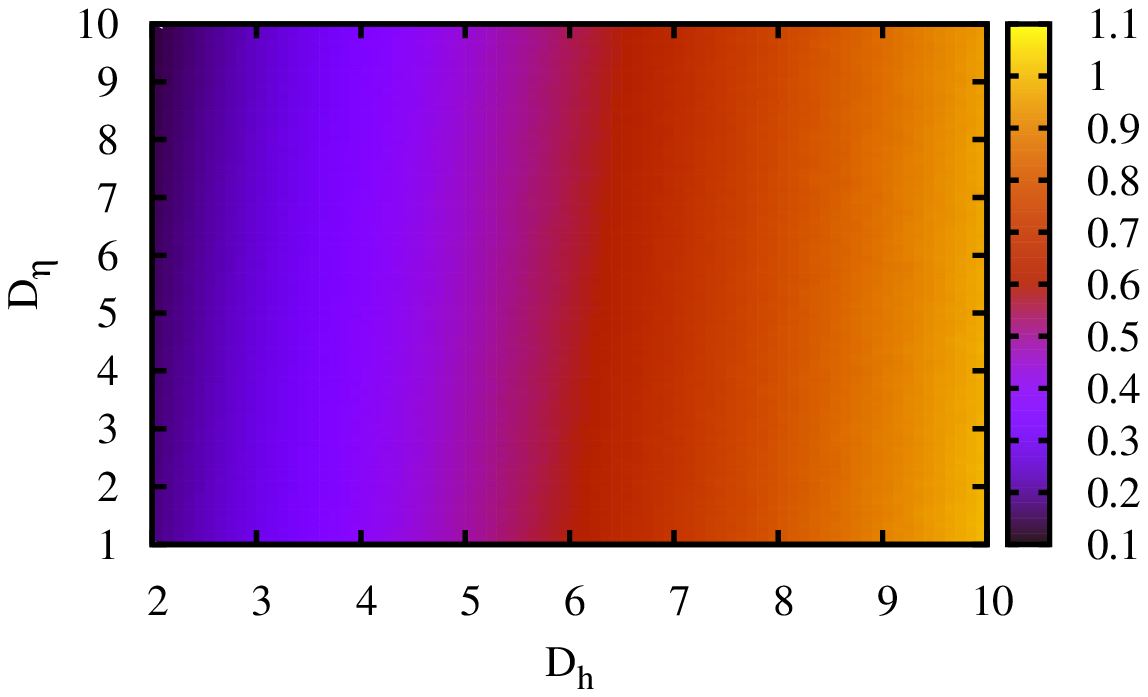}
 \caption{}
 \label{fig:power}
 \end{subfigure}
 \begin{subfigure}{0.48\linewidth}
 \includegraphics[width=\linewidth]{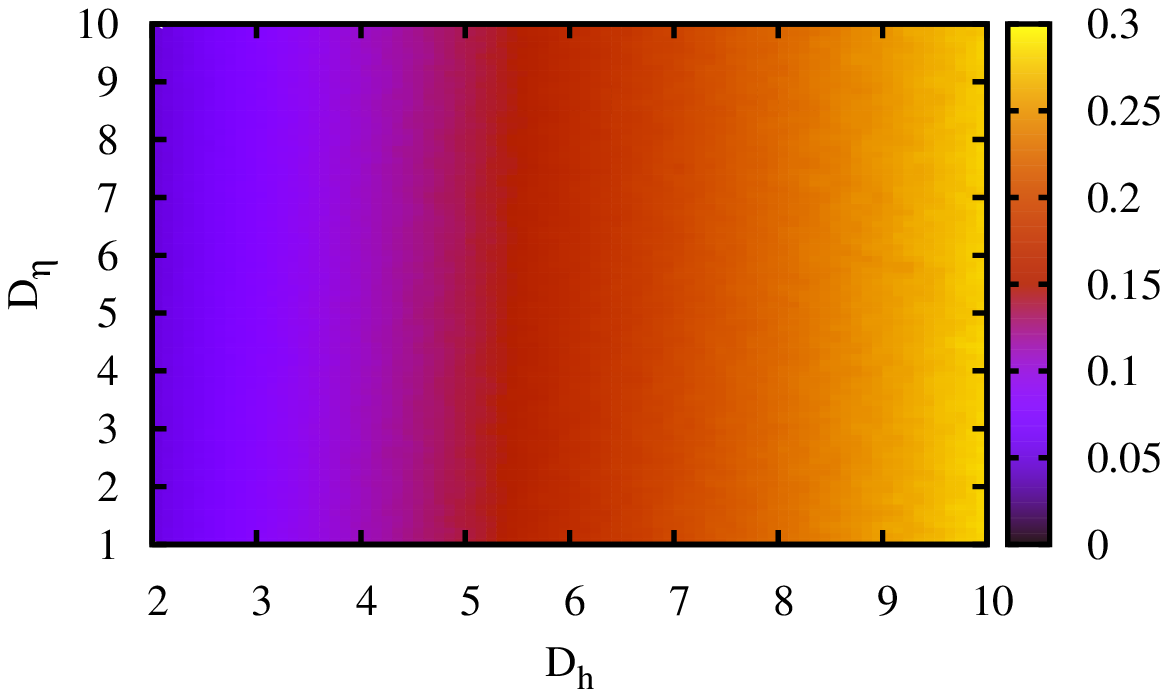}
 \caption{}
 \label{fig:tradeoff}
 \end{subfigure}
 \caption{(a) Phase plot of output power as a function of $D_\eta$ and $D_h$. (b) Phase plot of efficient power as a function of $D_\eta$ and $D_h$.}
\end{figure}

In figure \ref{fig:power}, we plot the map of power variation with the variations in active noise strength and thermal noise of hot bath. Comparing with figure \ref{fig:Dh_Eta_efficiency}, it is observed that the efficiency is generally smaller when the output power is higher, and vice versa. In figure \ref{fig:tradeoff}, we have shown the product of power and efficiency, $\eta \av W/\tau$, as a function of the same parameters. 
We find that the optimum region is obtained for $D_h>9$. Thus, for practical purposes, it would be preferable to allow the engine to work in this range of parameters.

\section{A simple application}
\label{sec:application}

\begin{figure}[!ht]
 \centering
 \resizebox{0.45\linewidth}{!}{
\begin{tikzpicture}%[thick,scale=1.2, every node/.style={transform shape}]
 \draw[line width=1.5, draw=black] (-1.2,-0.8) rectangle ++(6.5,3.5);
 \filldraw[red] (0,0.2) circle (5pt);
 \filldraw[black] (4.2,0.35) circle (5pt);
 \draw[line width=2,scale=1,domain=-1:1,smooth,variable=\x,blue] plot ({\x},{2*\x*\x});
 \draw[line width=2,scale=1,domain=3:5,smooth,variable=\x,blue] plot ({\x},{2*(\x-4)*(\x-4)});
 \draw[line width=1.2,decoration={aspect=0.2, segment length=2mm, amplitude=3mm,coil},decorate] (0,0.15)--(4.2,0.3);
 \node[] at (0.45,1.5) {\large $k(t)$};
 \node[] at (3.5,1.5) {\large $K$};
 \node[] at (2,1) {\large $K'$};
\end{tikzpicture}
}
\caption{Simple model for an engine attached to a flywheel}
\label{fig:flywheel}
\end{figure}
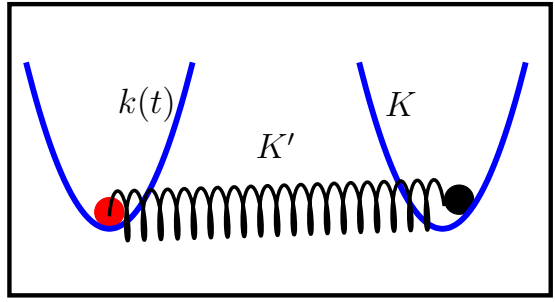

A macroscopic heat engine is often connected to a flywheel. For the microscopic engine that we are considering, we propose a simple model that can approximate the behaviour of such a flywheel. To do so, we connect the engine (particle 1) via a spring of stiffness constant $K'$ (acting as a soft shaft) to the microscopic ``flywheel'' (particle 2). The flywheel is another Brownian particle, that is in the same heat reservoir as the engine, but is trapped by a harmonic potential of constant stiffness $K$. This second trap models the pinning of the axle of the flywheel to a given point. The schematic diagram of this combined system has been provided in figure \ref{fig:flywheel}.

Let $x_1$ and $x_2$ be the respective displacements of particle 1 and particle 2 from the corresponding mean positions.
It is essentially a coupled oscillator system, and the equations of motion are given by
\begin{widetext}
\begin{alignat}{2}
&\begin{rcases}
 \gamma \dot{x}_1 = -(k_e(t)+K')x_1 + K'x_2 + \left(\sqrt{D_h}\right)\xi;\\
 \gamma \dot{x}_2 = K'x_1-(K+K')x_2+ \left(\sqrt{D_h}\right)\xi 
 \hspace{3cm}
 \end{rcases}&&\mbox{$0<t\le\tau/2$}\nonumber\\
 &\begin{rcases}
 \gamma \dot{x}_1 = -(k_c(t)+K')x_1 + K'x_2 + \left(\sqrt{D_c}\right)\xi + \left(\sqrt{D_\eta/\tau_\eta}\right)\eta\\
 \gamma \dot{x}_2 = K'x_1-(K+K')x_2+ \left(\sqrt{D_c}\right)\xi. 
 \end{rcases}&&\mbox{$\tau/2<t\le\tau$}. 
\end{alignat}
\end{widetext}

% 
%Correspondingly, the equations for the engine become
%
%
As before, the thermal noise $\xi$ is Gaussian and delta-correlated while the active noise $\eta$ is Gaussian and exponentially correlated.

\begin{figure}[!ht]
 \centering
 \includegraphics[width=0.6\linewidth]{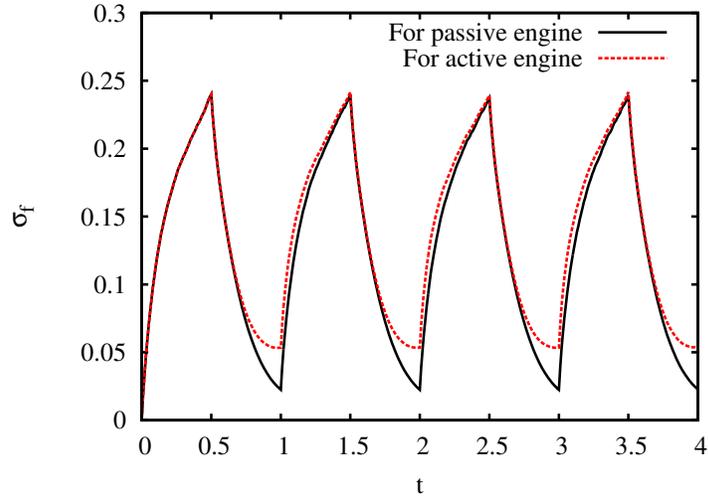}
 \caption{Time-dependence of the variance in the position of flywheel. Parameters used: $D_h=2,~D_\eta=10,~D_c=0.1,~\tau=1,~\tau_\eta=1,~K'=10,~K=0.1$.	}
 \label{fig:sigma_f}
\end{figure}

The combined system has been evolved numerically, and the variance $\sigma_f$ of the flywheel's position has been plotted as a function of time in figure \ref{fig:sigma_f} (red dashed line). For comparison, the corresponding curve for a flywheel attached to a passive engine has been shown (black solid line). We clearly observe that the flywheel attached to the active engine undergoes travels further from the mean position, especially when the engine is in its active state (compression cycle). Since the work done by the flywheel is proportional to the area under this curve, it means that greater work is done by the flywheel when the engine is active.

\section{Conclusions}

In this article, we have explored the thermodynamics of a microscopic heat engine formed of a single active particle, that undergoes a Stirling cycle. For the first cycle, we have obtained analytical results that agree very well with the results of our simulations. 
These results are true for the general case that includes the non-quasistatic regime of operation of the engine, and the combined effect of thermal and active noises.
We have then studied the engine in its steady state, i.e. when the position distribution varies periodically in time. We find that in a suitable parameter range the efficiency of the engine increases with increase in the active noise. This is a result that can be practically very useful while designing such engines. It is  observed that the active Stirling engine behaves like a passive one when the temperature difference between the thermal reservoirs is very high, and beyond a certain value of the temperature of the hot thermal bath, the efficiencies saturate to a limiting value. We have also shown the variation of the efficient power of the engine with the active noise and the thermal noise of the hot bath, and deduced the region where it is maximized. 
 A simple model that mimics the action of the engine on a flywheel has been provided, where it is observed that the flywheel does higher work when the engine is active. 
An extension of the concepts developed in this article to multiple active particles is under way. 
Furture studies may be conducted to shed light on the effect of anharmonicity on the properties of the engine. The corresponding problem for a quantum harmonic trap would also be interesting.

\section{Acknowledgements}

SL and AK thank DST-SERB, India for funding under research proposal sanction number ECR/2017/002607. PSP acknowledges the support from Foundational Questions Institute and Fetzer Franklin Fund, a donor advised fund of Silicon Valley Community Foundation under grant number FQXiRFP-1808. AS thanks University Grants Commission Faculty Recharge Program (UGCFRP), India for startup grant.

\appendix

\section{Derivation of efficiency for a quasistatic Stirling engine in absence of activity}

\label{sec:appendix}

In this section, we provide the derivation of the efficiency in the limit of $\tau\to \infty$ (see \cite{bec12_nature}). Note that the two isochoric steps are still instantaneous, and therefore only the isothermal steps are quasistatic.

The variances in this case become    $\sigma_{e,qs}=\lim_{t\to\infty}\sigma_e(t)=\frac{T_{h}}{k_{e}(t)}$, and $\sigma_{c,qs}=\lim_{t\to\infty}\sigma_c(t)=\frac{T_{c}}{k_{c}(t)}$.
 The work done in the cycle is given by
\begin{align}
 \langle W\rangle &= \frac{1}{2}\int_0^{\tau/2} \dot k_e(t) \sigma_{e,qs}(t) dt + \frac{1}{2}\int_{\tau/2}^\tau \dot k_c(t) \sigma_{c,qs}(t) dt\nonumber\\
 &= -\frac{T_{h}}{2}\ln\left(\frac{k_{max}}{k_{min}}\right) + \frac{T_{c}}{2}\ln\left(\frac{k_{max}}{k_{min}}\right).
 %&= -\frac{T_{h}}{2}\left(1-\frac{T_{c}}{T_{h}}\right)\ln\left(\frac{k_{max}}{k_{min}}\right).
 \label{W}
\end{align}
Note that no work is done in the isochoric steps, since the potential does not change during these steps. In order to compute the heat absorbed during the isothermal expansion, one must calculate the change in internal energy during this step. The mean heat absorbed is then given by $\langle Q_h\rangle = \langle W_h\rangle - \langle \Delta E_h\rangle$, where the subscript implies that during this step the particle is in contact with the hot bath at temperature $T_h$. 

Note that in the steady state, distributions at times $t=0^+$ and $t=\tau^-$ are given respectively by
\begin{align}
 P_0(x) &= \sqrt{\frac{k_{max}}{2\pi T_{h}}}\exp\left(\frac{- k_{max}~ x^2}{2T_{h}}\right).\nonumber\\
 P_\tau(x) &= \sqrt{\frac{k_{max}}{2\pi T_{c}}}\exp\left(\frac{- k_{max}~ x^2}{2T_{c}}\right).
 \label{P0}
\end{align}
%
% The position distribution at the end of the isothermal expansion and isothermal compression are given respectively by
% %
% \begin{align}
%  P_{\tau/2}(x) &= \sqrt{\frac{k_{min}}{2\pi T_{h}}}\exp\left(\frac{- k_{min}~ x^2}{2T_{h}}\right);\nonumber\\
%  P_\tau(x) &= \sqrt{\frac{k_{max}}{2\pi T_{c}}}\exp\left(\frac{- k_{max}~ x^2}{2T_{c}}\right).
% \end{align}
% %
It can be easily shown that there is no change in energy of the system during a quasistatic isothermal expansion. The only energy change in the first step must come from the relaxation of the final distribution of step 3 to the thermal distribution corresponding to the higher temperature in step 1. A sketch of a single cycle has been shown in figure \ref{fig:stirling}). 
The change in average energy is therefore
\begin{align}
 \left<\Delta E_h\right> &= \frac{1}{2}k_{max}(\left<x^2\right>_{0^+}-\left<x^2\right>_{\tau^-}) = \frac{1}{2}(T_h-T_c).
 \label{DeltaE}
\end{align}
Here the symbols $\left<\cdots\right>_{0^+}$ and $\left<\cdots\right>_{\tau^-}$ indicate that the averages have been taken over the distributions $P_0(x)$ and $P_{\tau}(x)$, respectively. Using Eqs. \eqref{W} and \eqref{DeltaE}, we finally arrive at the expression for efficiency:
\begin{align}
\eta_{stirling} &= \frac{\left<W\right>}{\left<W_h\right>-\left<\Delta E_h\right>} = \frac{\eta_{c}} {1+\eta_{c}\Big\{\ln\big[\frac{k_{max}}{k_{min}}\big]\Big\}^{-1}},
%\label{quasistatic}
\end{align}
where $\eta_c\equiv 1-T_c/T_h$ is the Carnot efficiency. Substituting the values of the parameters: $\tau=20$ (quasistatic limit), $D_h=2,~D_c=0.1$ and $k_0=10$ (Note that the corresponding temperatures are given by $T_h=D_h/2$ and $T_c=D_c/2$), the numerically obtained value is 0.38, which is in very good agreement with the theoretical value of 0.39. 

\section{Expression for $\sigma_c(t)$}
\label{sec:appendix2}

The expression for the variance $\sigma_c(t)$ during the compression arm of the first (transient) cycle is given by,
\begin{align}
 \sigma_c(t) &= \frac{D_h}{2\gamma^{3/2}}\left(\sqrt{\frac{\pi\tau}{k_0}}\right)e^{-\alpha_1(t)+k_0\tau/4\gamma}\left[\mbox{erf}\left(\sqrt{k_0\tau/\gamma}\right)-\mbox{erf}\left(\sqrt{k_0\tau/4\gamma}\right)\right] \nonumber\\
 &+ e^{-\alpha_1(t)}\int_{\tau/2}^t dt'~e^{\alpha_1(t')}\left[\frac{D_c}{\gamma^2}+\chi e^{-\alpha(t')/2-t'/\tau_\eta}\left\{\mbox{erfi}(\alpha_2(t'))-\mbox{erfi}(\alpha_2(\tau/2))\right\}\right],
\end{align}
where
\begin{align}
 \alpha_1(t) &= \frac{k_0(t^2-(\tau/2)^2)}{\gamma\tau}, \hspace{0.5cm}\alpha_2(t) = \frac{\gamma\tau/\tau_\eta+k_0 t}{\sqrt{2k_0\gamma\tau}},\nonumber\\
 \mbox{and} \hspace{1cm} \chi &= \frac{D_\eta}{\gamma^{3/2}\tau_\eta}\left(\sqrt{\frac{\pi\tau}{2k_0}}\right)\exp\left(\frac{k_0\tau}{8\gamma}-\frac{\gamma\tau}{2k_0{\tau_\eta}^2}\right).
\end{align}

% \begin{figure}[!ht]
%  \centering
%  \includegraphics[width=0.6\linewidth,height=0.4\linewidth]{phase_plot_Dc_Dh.png}
%  \caption{The plots show efficiency as a function of the bacterial activity. The increased efficiency for higher activity is apparent.}
% \end{figure}

		
%\bibliographystyle{apsrev4-1}
%\bibliography{ref}
\end{document}